\documentclass[prd,twocolumn,a4paper]{revtex4-1}
\usepackage{hyperref}
\usepackage{graphicx}
\usepackage{amsmath}
\usepackage{xcolor}
\usepackage{float}
\usepackage{amsfonts}
\usepackage{epsfig}
\usepackage{graphics}
\usepackage{color}
\usepackage{csquotes}

\newcommand{\beq}{\begin{equation}}
\newcommand{\eeq}{\end{equation}}
\newcommand{\bqa}{\begin{eqnarray}}
\newcommand{\eqa}{\end{eqnarray}}

\begin{document}
\title{Study of Differential Scattering Cross-section using Yukawa term of medium-modified Cornell potential}
\author{Siddhartha Solanki$^{a}$}
\email{siddharthasolanki2020@gmail.com}
\author{Manohar Lal$^{a}$} 
\email{manoharlalphd2019@gmail.com}
\author{Vineet Kumar Agotiya$^{a}$}
\email{agotiya81@gmail.com}
\affiliation{$^a$Department of Physics, Central University of Jharkhand, Ranchi, India, 835222}

\begin{abstract}
In the present work we have studied the differential scattering cross-section for ground states of charmonium and bottomonium in the frame work of the medium modified form of quark-antiquark potential and Born-approximation using the non-relativistic quantum chromo-dynamics approach. To reach this end, quasi-particle (QP) Debye mass depending upon baryonic chemical potential ($\mu_{b}$) and temperature has been employed and hence the variation of differential scattering cross-section with baryonic chemical potential and temperature at fixed value of the scattering angle ($\theta$=$90^o$) has been studied. The variation of differential scattering cross-section with scattering angle $\theta$ (in degree) at fixed temperature and baryonic chemical potential has also been studied. We have also studied the effect of impact parameter and transverse momentum on differential scattering cross-section at $\theta$=$90^o$. \\

{\bf Keywords} : Differential Cross-section, Scattering theory, Strongly Coupled Plasma, Heavy Quark Potential, baryonic chemical potential, Born-approximation, Scattering amplitude, quasi-particle Debye mass.
\end{abstract}

\maketitle

\section{Introduction}
Quantum Chromo-dynamics (QCD) is one among the most important theories which well describes the strong interaction occurring at the subatomic level. Since the cross-section is an important key that pave a way of communication between the real world of the experiment and idealized theoretical models. In the high energy physics the term cross-section is used to specify the interaction of elementary particle quantitatively. Cross-section may also be thought of as the area within which the reaction among the elementary particle takes place. Theoretical predictions for the cross-section of the oppositely lepton pairs in p-p collisions are accurate up to the next to leading order (NLO) and next to next leading order (NNLO) in electroweak and perturbative quantum chromo-dynamics (PQCD)~\cite{1,2,3,4} respectively. Precise measurements of the differential cross-section at LHC for the Drell Yan process (end up with conformal test for the standard model in the perturbative regime) is an important test for the standard model in the perturbative frame. ATLAS~\cite{5,6,7} and CMS~\cite{8,9,10} recently measured the single and double cross-section.\\ 
Inclusive quarkonia production in pp collisions, in pp collision at $\sqrt{s}$=$5.02$ TeV. Both the perturbative and non-perturbative fact of QCD can be studied easily in high-energy hadronic collision by considering the quarkonia production~\cite{11,12}. The main consequences of scattering process in hadronic collision is to produce quarkonia. In such process the momentum transfer should be two times the mass of heavy quark and hence it can described under perturbative calculations. But on the other hand, the binding energy of quarkonia comes under the non-perturbative process as it account for large distance scales and soft momentum scales. Various properties of the Quark-gluon plasma (QGP) in nucleus-nucleus collision at different energy scales and that of cold matter nuclear effect appearing in A-A collision could be investigated by the quarkonium production measurement~\cite{12,13}. The quarkonium production can be described by various approaches.
One of the most important countering problem in the QCD is to fully understand the production mechanism of the heavy quarkonia since after the discovery of $J/\psi$ in 1973. According the colored singlet model the quarkonium states are color-less and they possesses the same $J^{pc}$ quantum number~\cite{14,15,16,17,18,19,20,21,22,23}.\\
Production of $J/\psi$ and $\psi(2s)$ cross-section at high $P_T$ was underestimated by the leading order (LO) calculation in colored singlet model by one order of magnitude~\cite{24} and this problem was overcome by considering the next to leading order (NLO) correction but this result would not be still able to reduce the gap between color singlet model (CSM) and experimental measurements~\cite{25,26,27}. Thereafter the non-relativistic QCD model came into existence, which includes both color singlet and color octet studies~\cite{28,29}, describing  production cross-section at all $P_T$ values but fails to explain the polarization~\cite{30,31,32,33,34,35,36,37,38,39,40,41,42,43,44,45}. Finally, studies~\cite{46,47,48,49,50,51} solved this countering problem via production of pairs of quarkonia, as the cross-section could be easily interpreted. This quarkonium pair production forbid the feed down of excited C-even states which are very crucial in the single quarkonia production. This typically makes the interpretation of polarization very difficult and hence to compare the data. The double parton process has a significant role for understanding the new physics e.g multi-jets and gave a pave for the transverse momenta of partons. Several new physics phenomenon have been studied by keeping in view of double parton process such as $4$-jets by AFS~\cite{52}, UA2~\cite{53}, CDF~\cite{54} and ATLAS~\cite{55} collaborations. $\gamma$+$3$ jets by the CDF~\cite{56} and $D_o$~\cite{57,58} collaborations $W+2jets$~\cite{59} and $\Upsilon$+$\Upsilon$~\cite{60} by CMS collaborations $J/\psi$+W~\cite{61}, Z+open charm~\cite{62}, and $\Upsilon$+open charm~\cite{63} by the $LHC_b$ collaborations. Quarkonia pairs are independently produced by different partonic interaction in the frame work of double parton process could by estimated by the formula~\cite{64,65,66}.\\
In this present paper, we studied the differential scattering cross-section for the ground state of Quarkonia (i.e, $J/\psi$ and $\Upsilon$), in the presence of temperature, baryonic chemical potential ($\mu_{b}$) and the scattering angle ($\theta$). To carry out this, we use the Born approximation for calculating the Scattering amplitude and differential scattering cross-section. For calculating the differential scattering cross-section,the potential we consider is the medium-modified form of Cornell potential (Here we take only the Yukawa term and neglect the other terms of the potential). differential scattering cross-section mainly define the probability of finding the particle in a certain area and hence scattering angle $\theta$ plays a major role while studying the quarkonium production.\\ 
The manuscript organized in the following manner. In the section-II, we provide a outline for the quark-antiquark potential. Section-III, deals with the Debye mass depending upon temperature and baryonic chemical potential. Section-IV, Formulation of differential scattering cross-section using Non-relativistic limit of Quantum field theory. In section-V, we briefly discussed about the result and Conclusions of this present work.
\begin{figure*}
\vspace{2mm}
\includegraphics[scale=.60]{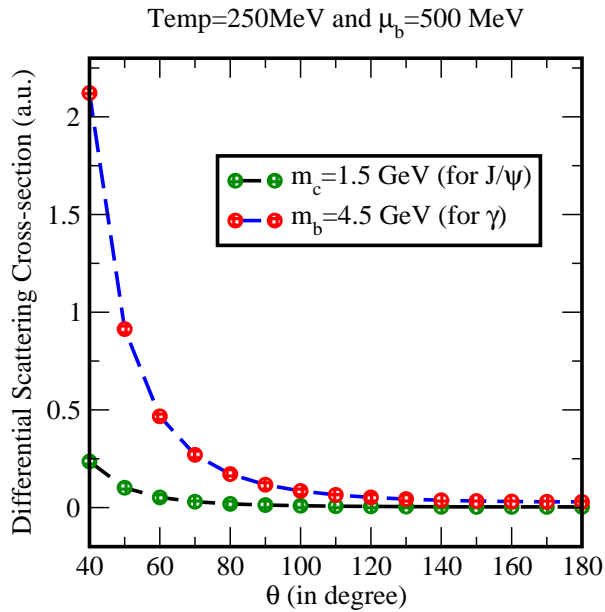}
\vspace{2cm}
\caption{The variation of differential scattering cross-section as a function of theta after considering the values of charmonium and bottomonium masses at constant temperature and $\mu_{b}$.}
\label{fig.1}
\vspace{2cm}
\end{figure*}

\section{Medium-modified form of Cornell potential}
Since from~\cite{67}, it has been seen that there is also non-perturbative calculations at deconfinement temperature instead of the perturbative and ideal gas behavior according to thermodynamical studies of the QCD. Following~\cite{67} one cannot drops the string tension arising between the quark-antiquark pairs above the critical temperature $T_{c}$. Medium modification to the quark-antiquark potential provide a reliable way to study the fate of the quarkonia. Here the quark-antiquark potential has been corrected which embodied the medium effect~\cite{68}. In~\cite{69,70}, the authors assume that the string is melting, keeping in view that there is phase transition from the hadron matter to the quark gluon plasma. Accordingly, the potential has been modified to study the deconfined state of matter. Also at vanishing baryon density there is cross-over rather than a phase transition. These indications comes from the lattice studies. Bound state solutions of both relativistic and non-relativistic wave has attracted more intention from the last decades.\\ 
The energy of these bound states is negative due to the fact that the energy of the quark gluon plasma is less than the potential energy~\cite{71}. Schrodinger equation accounts for the non-relativistic wave whereas for the relativistic wave Klein-Gordan and Dirac equation are of utmost important~\cite{72,73,74,75,76,77,78,79}. There are various potential has been used to study the quarkonia bound states like Hulthen, Poschl Teller, Eckart and coulomb potential etc. and these are studied using special techniques AEIM, SUSYQM and NU method~\cite{80,81,82,83,84,85,86,87,88,89,90} etc. In the present work, we preferred to work with the Cornell potential which has both coulombic as well as string part~\cite{91,92}, and here we take only the coulombic part of the said potential. This is because of the fact that the mass of quarkonia $m_{Q} \geq \lambda_{QCD}$, small velocity of the bound states push to understand these phenomenon in terms of non-relativistic potential model.
\begin{figure*}
\vspace{2mm}
\includegraphics[scale=.55]{fig2a.eps}
\hspace{5mm}
\includegraphics[scale=.55]{fig2b.eps}
\vspace{2cm}
\caption{The variation of $J/\psi$ differential scattering cross-section as a function of $\mu_{b}$ at different values of temperature (in left panel) and as a function of $T/T_{c}$ at different values of baryonic chemical potential (in right panel).}
\label{fig.2}
\vspace{2cm}
\end{figure*}

In case of finite temperature QCD, we employ the ansatz that the medium modification enter in the Fourier transform of heavy quark potential V(k) as~\cite{93}

\begin{equation}
\label{eq1}
\tilde{V(k)}=\frac{V(k)}{\varepsilon(k)}
\end{equation}.

where $\varepsilon(k)$ is dielectric permittivity which is obtain from the static limit of the longitudinal part of the gluon self energy.

\begin{equation}
\label{eq2}
\varepsilon(k)=\left(1+\frac{\pi_L(0,k,T)}{k^2}\right)\equiv\left(1+\frac{m^2_D(T,\mu_{b})}{k^2}\right)
\end{equation}.

V(k) is the Fourier transform of the Cornell potential given as,

\begin{equation}
\label{eq3}
\mbox{\boldmath$V$}(k)=-\sqrt{\frac{2}{\pi}}\frac{\alpha}{k^2}-\frac{4\sigma}{\sqrt{2\pi}k^4}
\end{equation}.

substituting the value of Eq.(\ref{eq2}) and Eq.(\ref{eq3}) in equation Eq.(\ref{eq1}), and solving using inverse Fourier transform, we get the medium modified potential depending upon 'r'~\cite{94,95,96,97,98}.

\begin{multline}
\label{eq4}
\mbox{\boldmath$V$}(r,T,\mu_{b})=\left(\frac{2\sigma}{m^2_D(T,\mu_{b})}-\alpha\right)\frac{exp(-m_D(T,\mu_{b})r)}{r}\\-\frac{2\sigma}{m^2_D(T,\mu_{b})r}+\frac{2\sigma}{m_D(T,\mu_{b})}-\alpha m_D(T,\mu_{b})
\end{multline}.

\section{The Debye mass with baryonic-chemical potential from a quasi-particle picture of hot QCD}
In studies of the quantum mechanical properties of the quarkonia, Debye mass has played a significant role. Generalization of the Debye mass has been made from the quantum electrodynamics (QED) to QCD because of the non-abelian nature of the QCD. In QCD, the Debye mass obtained is gauge invariant and non-perturbative whereas the temperature dependent Leading order Debye mass is perturbative and is known from a long time ago~\cite{99,100}. The Debye mass can also be defined as the pole static quark propagator~\cite{101} instead of limit $p {\longrightarrow} 0$ in the gluon self energy. Authors in~\cite{101,102} also calculated Debye mass for the NLO in QCD using Polyakov loop correlator matching with HTL result.\\
Several studies has been devoted to include all the interaction present in the Hot QCD equation of states (EoS) in terms of the quasi-partons. Some of them include effective mass models, effective mass models with Polyakov loop, PNJL, NJL model and effective fugacity model~\cite{103,104,104,106,107,109}.\\
To understand the non-ideal behavior of the Quark gluon plasma near the cross over region, quasi particle model has played eminent role. The interacting system of massless quarks and gluons considered as the massive system in Quasi-Particle (QP) model~\cite{109}. In our present calculation we use the quasi particle model to study the quarkonia properties. All the interaction effects could be related to the $Z_{q,g}$ term in the distribution function of the quasi partons.
\begin{figure*}
\vspace{2mm}
\includegraphics[scale=.55]{fig3a.eps}
\hspace{5mm}
\includegraphics[scale=.55]{fig3b.eps}
\vspace{2cm}
\caption{The variation of $\Upsilon$ differential scattering cross-section as a function of $\mu_{b}$ at different values of temperature (in left panel) and as a function of $T/T_{c}$ at different values of baryonic chemical potential (in right panel).}
\label{fig.3}
\vspace{2cm}
\end{figure*}

In our calculation, we use the Debye mass $m_{D}$ for full QCD case is,
\begin{eqnarray}
\label{eq5}
m^2_D\left(T \right) &=& g^2(T) T^2 \bigg[
\bigg(\frac{N_c}{3}\times\frac{6 PolyLog[2,z_g]}{\pi^2}\bigg)\nonumber\\&&
+{\bigg(\frac{\hat{N_f}}{6}\times\frac{-12 PolyLog[2,-z_q]}{\pi^2}\bigg)\bigg]}
\end{eqnarray}
and 
\begin{eqnarray}
\label{eq6}
\hat{N_f} &=& \bigg(N_f +\frac{3}{\pi^2}\sum\frac{\mu_{q}^2}{T^2}\bigg)
\end{eqnarray}
and we also know that, the quark-chemical potential is equal to,
\begin{eqnarray}
\label{eq7}
\mu_{q} &=& \frac{\mu_{b}}{3}
\end{eqnarray}
Where, $(\mu_{q})$ defined the quark-chemical potential and $(\mu_{b})$ is baryonic chemical potential. Introducing the value of $\hat{N_f}$ in the Eq.(\ref{eq5}), we get the full expression of quasi particle Debye mass in terms of temperature and baryonic chemical potential.
\begin{equation}
\label{eq8}
m^2_D\left(T,\mu_{b} \right)=T^2\left\{\bigg[\frac{N_c}{3} Q^2_g\bigg]+\bigg[\frac{N_f}{6}+\frac{1}{2\pi^2}\bigg(\frac{\mu_{b}^2}{9 T^2}\bigg)\bigg] Q^2_q\right\} 
\end{equation}
Here, $g(T)$ is the QCD running coupling constant, $N_{c}=3$ is the number of color and $N_{f}$ is the number of flavor, the function $PolyLog[2,z]$ having form, $PolyLog[2,z]=\sum_{k=1}^{\infty} \frac{z^k}{k^2}$ and $z_g$ is the quasi-gluon effective fugacity and $z_q$ is quasi-quark effective fugacity. 

\begin{eqnarray}
\label{eq9}
f_{g,q}=\frac{z_{g,q}exp(-\beta p)}{\left (1\pm z_{g,q}exp(-\beta p)  \right )}
\end{eqnarray}

These distribution functions are isotropic in nature. These fugacities have been introduced the all interaction effects present within the 
baryonic chemical potential. Both $z_g$ and $z_q$ have a very complicated temperature dependence and asymptotically reach to the ideal 
value unity~\cite{110}. The temperature dependence $z_g$ and $z_q$ fits well to the form given below,
\begin{equation}
\label{eq10}
z_{g,q}= a_{q,g}\exp\bigg(-\frac{b_{g,q}}{x^2}-\frac{c_{g,q}}{x^4}-\frac{d_{g,q}}{x^6}\bigg)
\end{equation}

(Here $x=T/T_c$ and $a$, $b$ and $c$ and $d$ are fitting parameters), for both EOS1 and EOS2. Here, EoS1 is the $O(g^5)$ hot QCD and EoS2 is the $O(g^6\ln(1/g)$ hot QCD EoS in the quasi-particle description~\cite{94,107} respectively. Where, $Q_g$ and $Q_q$ are the effective charges given by the equations:
\begin{eqnarray}
\label{eq11}
 Q^2_g&=&g^2 (T) \frac{6 PolyLog[2,z_g]}{\pi^2}\nonumber\\
 Q^2_q&=&g^2 (T)  \frac{-12 PolyLog[2,-z_q]}{\pi^2}
\end{eqnarray}
In our present analysis,the temperature and baryonic chemical potential dependent quasi-particle Debye mass, $m_D^{QP}$ in full QCD with $N_f=3$ has been employed to study the differential scattering cross-section of the ground states of quarkonia.

\section{Formulation of differential scattering cross-section using Non-relativistic limit of Quantum field theory (QFT)}
In the non-relativistic limit, the QFT equation for the S-matrix reduce to the Lippmann-Schwinger equation for the scattering amplitude. The Lippmann-Schwinger equation is equivalent to the Schrodinger equation. In non-relativistic quantum mechanics the first order Born approximation of the elastic scattering amplitude is given by the Fourier transform of the potential. Correspondingly, the potential is given as inverse Fourier transform of the Scattering amplitude.\\                                 
In classical scattering theory, the essential countering problem is: (a) to measure the impact parameter, (b) to calculate the scattering angle. But in the Quantum scattering theory, the solution to the Schrodinger equation pave a good way to understand the scattering process for the proper wave function. Here quantum description of scattering of non-relativistic particles of mass $m_{1}$ and $m_{2}$ has been considered. For simplicity, we consider the case of elastic scattering. The interacting potential between particles is supposed to be time independent, and obviously time independent Schrodinger equation with the wave function has been used to obtained the scattering amplitude formula, for the calculation of differential scattering cross-section.
\begin{figure*}
\vspace{2mm}
\includegraphics[scale=.70]{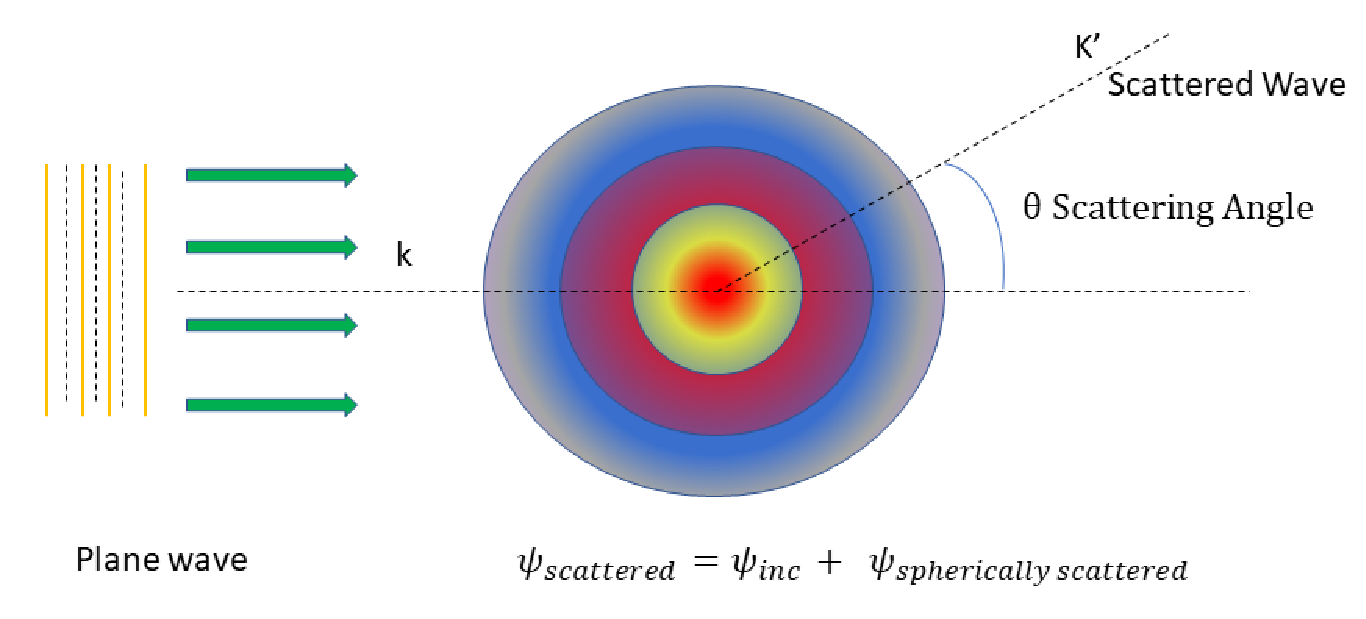}
\vspace{2cm}
\caption{This figure shows the interaction of scattered wave and plane wave with the potential V(r).}
\label{fig.4}
\vspace{2cm}
\end{figure*}

\begin{equation}
\label{eq12}
\left [ -\frac{\hbar^{2}}{2m_1}\bigtriangledown _{1}^{2}-\frac{\hbar^{2}}{2m_2}\bigtriangledown _{2}^{2}+ V(\vec{r_{1}} , \vec{r_{2}}) \right ]\psi(\vec{r_{1}} , \vec{r_{2}})=E_{T}\psi(\vec{r_{1}} , \vec{r_{2}})
\end{equation}

Where, $E_{T}$ is the total energy of the system, and this body problem can be reduced into a one-body problem then the Schrodinger equation will be,

\begin{equation}
\label{eq13}
\left [ -\frac{\hbar^{2}}{2\mu}\bigtriangledown^{2}+ V(\vec{r}) \right ]\psi(\vec{r})=E_{T}\psi(\vec{r})
\end{equation}

Now, we have to find wave function by solving Eq.(\ref{eq13}). And this is obtained by complex calculations using green function as following,

\begin{equation}
\label{eq14}
\psi (\vec{r})=\Phi_{inc}(\vec{r})-\frac{\mu}{2\pi\hbar^{2}}\int \frac{e^{ik|\vec{r}-\vec{{r}'}|}}{|\vec{r}-\vec{{r}'}|}V(\vec{{r}'})\psi(\vec{{r}'})d^{3}{r}' 
\end{equation}

Where, $\psi(\vec{r})$ is the wave function after scattering. To find scattering amplitude we will apply asymptotic limit on Eq.(\ref{eq14}) and will compare this wave function to the wave function in asymptotic limit that is discussed below for asymptotic limit is,

\begin{equation}
\label{eq15}
\psi(r,\theta)\cong A\left\{ e^{ikz} + f(\theta,\Phi) \frac{e^{ikr}}{r} \right\} 
\end{equation} 

The wave function in asymptotic limit after scattering will contain a unscattered plane wave plus a scattered spherical wave. We consider $A=1$ because it don't contribute in $ d\sigma/d\pi$. 
\begin{figure*}
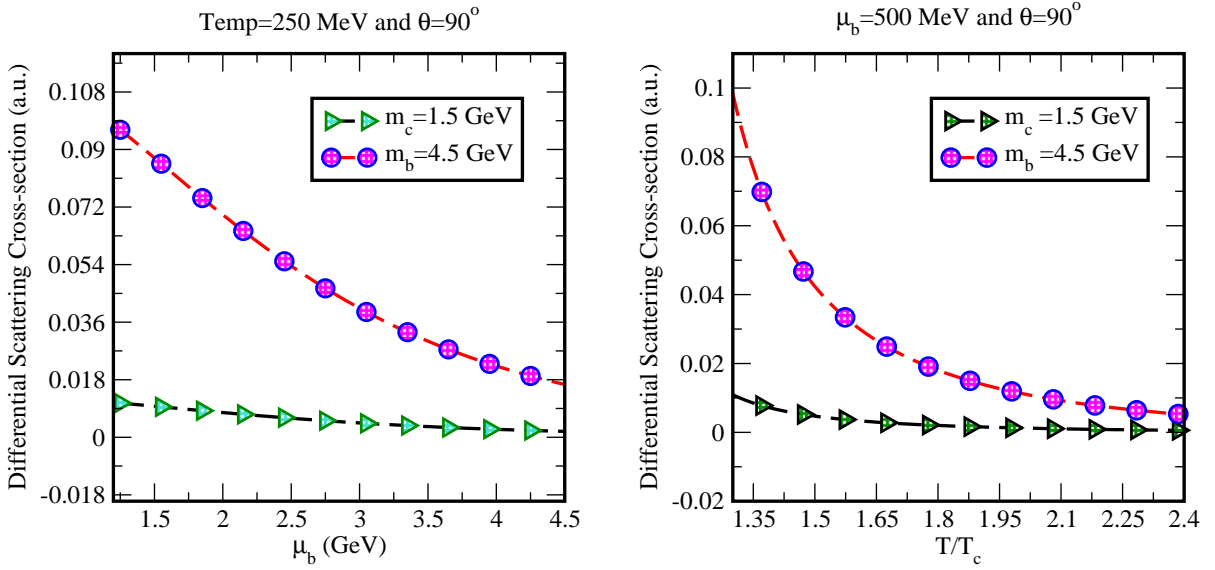

\vspace{2mm}
\includegraphics[scale=.55]{fig5a.eps}
\hspace{5mm}
\includegraphics[scale=.55]{fig5b.eps}
\vspace{2cm}
\caption{The variation of differential scattering cross-section as a function of $\mu_{b}$ at fixed values of temperature and theta (in left panel) and as a function of $T/T_{c}$ at fixed value of $\mu_{b}$ and theta (in right panel) after considering the values of charmonium and bottomonium masses.}
\label{fig.5}
\vspace{2cm}
\end{figure*}

\begin{equation}
\label{eq16}
k|\vec{r}-\vec{{r}'}|=kr- \vec{k}\vec{{r}'}
\end{equation}

\begin{equation}
\label{eq17}
\frac{1}{|\vec{r}-\vec{{r}'}|}=\frac{1}{r}
\end{equation}

\begin{equation}
\label{eq18}
\psi (\vec{r})=\Phi_{inc}(\vec{r})-\frac{\mu}{2\pi\hbar^{2}}\int \frac{e^{i(kr-\vec{k}\vec{{r}'})}}{r}V(\vec{{r}'})\psi (\vec{{r}'})d^{3}{{r}'}
\end{equation}

\begin{equation}
\label{eq19}
\psi (\vec{r})=\Phi_{inc}(\vec{r})-\frac{\mu}{2\pi\hbar^{2}}\frac{e^{ikr}}{r}\int e^{-ikr'} V(\vec{{r}'})\psi (\vec{{r}'})d^{3}{{r}'}
\end{equation}

Now, compare it with,
\begin{figure*}
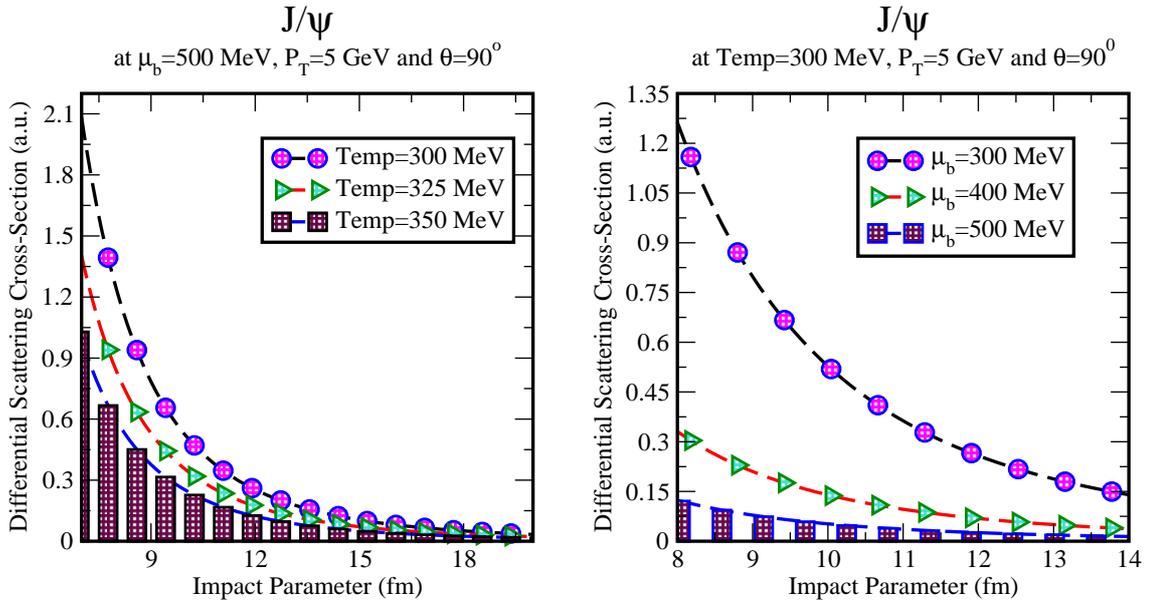

\vspace{2mm}
\includegraphics[scale=.55]{fig6a.eps}
\hspace{5mm}
\includegraphics[scale=.55]{fig6b.eps}
\vspace{2cm}
\caption{The variation of $J/\psi$ differential scattering cross-section as a function of impact parameter at different values of temperature (in left panel) and at different values of baryonic chemical potential (in right panel).}
\label{fig.6}
\vspace{2cm}
\end{figure*}

\begin{equation}
\label{eq20}
\psi (\vec{r})=\Phi_{inc}(\vec{r})+f(\theta ,\Phi )
\end{equation}

So, Scattering amplitude is,

\begin{equation}
\label{eq21}
f(\theta,\Phi)=-\frac{\mu}{2\pi\hbar^{2}}\int e^{-i\vec{k}\vec{{r}'}}V(\vec{{r}'})\psi (\vec{{r}'})d^{3}{r}'
\end{equation}

Now, using first order Born Approximation then the above expression is,

\begin{equation}
\label{eq22}
f(\theta,\Phi)=-\frac{\mu}{2\pi\hbar^{2}}\int e^{i\vec{q}\vec{{r}'}}V(\vec{{r}'})d^{3}{r}'
\end{equation}
Now, we consider the spherically symmetric potential,
\begin{figure*}
\vspace{2mm}
\includegraphics[scale=.55]{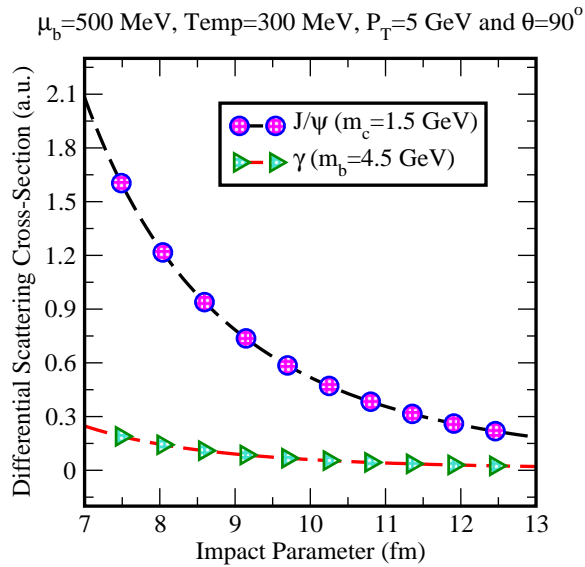}
\vspace{2cm}
\caption{The variation of differential scattering cross-section as a function of impact parameter after considering the values of charmonium and bottomonium masses.}
\label{fig.7}
\vspace{2cm}
\end{figure*}

\begin{equation}
\label{eq23}
f(\theta,\Phi)=-\frac{\mu}{2\pi\hbar^{2}}\int e^{i\vec{q}\vec{{r}'}}V(r)r^{2}sin\theta dr d\theta d\Phi   
\end{equation}

This is the modified form of Born approximation. But the simplest form of spherical symmetry amplitude after complete solution we get,

\begin{equation}
\label{eq24}
f(\theta,\Phi)\cong -\frac{2m_{{Q\bar{Q}}}}{q\hbar^{2}}\int_{0}^{\infty }rV(r)sinqrdr
\end{equation}

The angular dependence of $f$ is carried by $q$, 
\begin{figure*}
\vspace{2mm}
\includegraphics[scale=.55]{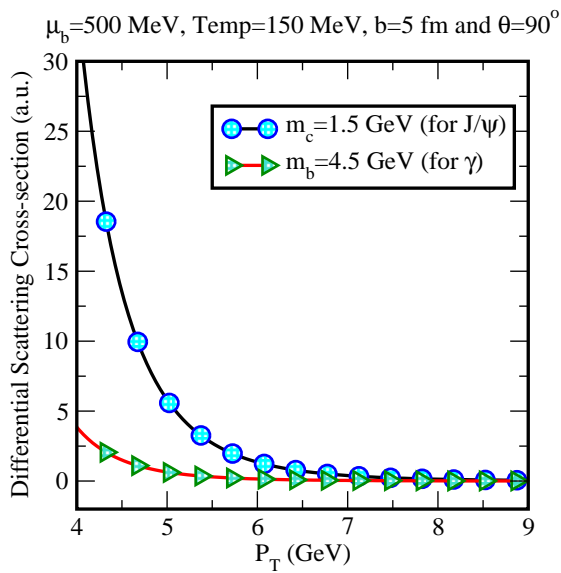}
\vspace{2cm}
\caption{Variation of the differential scattering cross-section with the transverse momentum ($P_{T}$) at $\mu_{b}$=500 MeV, $\theta$=$90^o$,and impact parameter b=5 fm and temperature T=150 MeV.}
\label{fig.8}
\vspace{2cm}
\end{figure*}
\begin{equation}
\label{eq25}
q=2ksin\frac{\theta}{2}
\end{equation}

We consider the r-dependence of the medium modified Cornell potential Eq.(\ref{eq4}), for the calculation of differential scattering cross-section, we consider only Yukawa term and neglect the constant term of the equation Eq.(\ref{eq4}) then we get,
 
\begin{equation}
\label{eq26}
V(r)=\left[ \frac{2\sigma}{m_{D}^{2}}-\alpha \right]\frac{exp(-m_{D}r)}{r}-\frac{2\sigma}{m_{D}^{2}r}
\end{equation}

Now, we used Eq.(\ref{eq24}) (scattering amplitude formula) for the above potential Eq.(\ref{eq26}) to calculate the differential scattering cross-section, and hence the result of differential scattering cross-section is shown below,
\begin{multline}
\label{eq27}
\frac{d\sigma}{d\Omega}=\left| f(\theta, \mu_{b}) \right|^{2}=\left[ \frac{2\sigma}{m_{D}^{2}}-\alpha \right]^{2}\left[ \frac{m_{q}}{m_{D}^{2}+(10sin\frac{\theta}{2})^{2}} \right]^{2}\\+\left[ \frac{2m_{q}\sigma}{m_{D}^{2}(10sin\frac{\theta}{2})^{2}} \right]^{2}
\end{multline}
We also calculate the differential scattering cross-section in terms of impact parameter and momentum transfer which is described below,
\begin{equation}
\label{eq28}
b=\frac{a}{2}cot\frac{\theta }{2}
\end{equation}
After considering the semi-classical consideration, for H-atom like problem ($z_{1}e$=$z_{2}e$ $\approx$ $1$) the value of a is,
\begin{equation}
\label{eq29}
a=\frac{z_{1}z_{2}e^{2}}{mv^{2}}
\end{equation}
So, after considering the Eq.(\ref{eq29}) the Eq.(\ref{eq28}) is changes into the Eq.(\ref{eq30}) i.e,
\begin{equation}
\label{eq30}
sin\frac{\theta }{2}=\frac{m_{q}cos\frac{\theta }{2}}{2P_{T}^{2}sin\frac{\theta }{2}}
\end{equation}
Now, substituting the value of Eq.(\ref{eq30}) into the Eq.(\ref{eq27}), then the differential scattering cross-section is converted into the Eq.(\ref{eq31}) in terms of impact parameter and momentum transfer.
\begin{multline}
\label{eq31}
\frac{d\sigma }{d\Omega }=|f(\theta, \mu_{b})|^{2}=\left [ \frac{2\sigma }{m_{D}^{2}}-\alpha  \right ]^{2}\left \{ \frac{m_{q}}{m_{D}^{2}+\left [ \frac{10m_{q}cos\frac{\theta }{2}}{2bP_{T}^{2}} \right ]^{2}} \right \}^{2}\\+\left \{ \frac{2m_{q}\sigma }{m_{D}^{2}\left [ \frac{10m_{q}cos\frac{\theta }{2}}{2bP_{T}^{2}} \right ]^{2}} \right \}^{2}
\end{multline}

\section{Results and conclusions}
In the present paper, we have studied the differential scattering cross-section for quarkonium ground states i.e, charmonium (J/$\psi$ state) and bottomonium ($\Upsilon$ state) using Non-relativistic limit of Quantum field theory (QFT). Figure~\ref{fig.1}, shows the variation of differential scattering cross-section as a function of theta ($\theta$) for different masses of the quarkonia (for J/$\psi$=1.5 GeV and for $\Upsilon$=4.5 GeV). It has been seen from figure~\ref{fig.1} that the variation of differential scattering cross-section for $\Upsilon$ is greater than that of J/$\psi$ at T=250 MeV and $\mu_{b}$=500 MeV. It has been also observed that if the value of $\theta$ increases, the separation between the differential scattering cross-section of $\Upsilon$ and J/$\psi$ is decreases. This is because of the fact, with increase in the values of $\theta$, the probability of finding the particle (J/$\psi$ and $\Upsilon$) decreases.\\
Figure~\ref{fig.2} and Figure~\ref{fig.3}, shows that the variation of differential scattering cross-section as a function of $\mu_{b}$ (in left panel) at different values of temperature and as a function of $T/T_{c}$ (in right panel) at different values of $\mu_{b}$ at fixed value of $\theta$=$90^o$ for the J/$\psi$ and for $\Upsilon$ respectively. It has been seen that the differential scattering cross-section as a function of $\mu_{b}$ and temperature is decreases exponentially. If we increase the values of temperature (in left panel), then variation of differential scattering cross-section is also decreases. With increases the values of baryonic chemical potential $\mu_{b}$ (in right panel), the variation of differential scattering cross-section is decreases but the effect of $\mu_{b}$ is smaller as compared to the temperature which can be seen from the right panel of figure~\ref{fig.2} and figure~\ref{fig.3} respectively.\\
Figure~\ref{fig.4}, shows the interaction of plane wave with the potential V(r), after interaction, the plane wave scattered spherically. The interacting potential between particles is considered as time independent potential given in Eq.(\ref{eq26}) and here we use time independent Schrodinger equation to calculate the differential scattering cross-section expression which is given in Eq.(\ref{eq27}).\\
Whereas, figure~\ref{fig.5}, shows the variation of differential scattering cross-section as a function of baryonic chemical potential at fixed value of $\theta=90^o$ and temperature (T=250 MeV) (in left panel) and in right panel as a function of $T/T_c$ for the fixed value of $\mu_{b}$ and $\theta$ (i.e., $\mu_{b}$=500 MeV and $\theta=90^o$) and for mass $m_{J/\psi}$=1.5 GeV and $m_{\Upsilon}$=4.5 GeV respectively. It has been clearly seen from figure~\ref{fig.5} that the variation of differential scattering cross-section of $\Upsilon$ is greater than in comparison to $J/\psi$ because mass of $\Upsilon$ is greater than as compared to J/$\psi$.\\
Figure~\ref{fig.6} shows the variation of differential scattering cross-section for the fixed value of transverse momentum $P_T$=5 GeV and $\theta$=$90^o$ as a function of impact parameter at different values of temperature and fixed baryonic chemical potential $\mu_{b}$=500 MeV (in left panel) and at different values of the baryonic chemical potential and fixed temperature T=300 MeV (in right panel). Also, differential cross section of the quarkonium production is decreases with impact parameter. In case of impact parameter, the variation of differential scattering cross-section is also decreases with the increases of temperature and baryonic chemical potential.\\ 
Figure~\ref{fig.7} shows the variation of differential scattering cross-section for the fixed values of the temperature (T=300 MeV), baryonic chemical potential ($\mu_{b}$=500 MeV), transverse momentum ($P_T$=5 GeV) and $\theta$=$90^o$ with impact parameter for charmonium ($m_{J/\psi}$=1.5 GeV) and bottomonium ($m_{\Upsilon}$=4.5 GeV) masses. It has been clearly seen from figure~\ref{fig.7} that the variation of differential scattering cross-section of J/$\psi$ is greater than in comparison to $\Upsilon$ with respect to impact parameter.\\
Finally, figure~\ref{fig.8} shows that how the differential scattering cross-section varies with the transverse momentum at $\mu_{b}$=500 MeV, $\theta$=$90^o$, $b$=5 fm and $T$=150 MeV. It has been also observed that there is strong decrease in the differential cross-section with the transverse momentum for higher masses, and same behavior of deferential scattering cross-section is observed like impact parameter.\\ Usually the scales one encounters are $P_{T}$, $m_{Q}$, $m_{Q}\lambda$ and $m_{Q}\lambda^{2}$, where, $P_{T}$, $m_{Q}$ and $\lambda$ are the transverse momentum, heavy quark mass and heavy quark-anti-quark pair relative velocity in the quarkonium rest frame ( $\lambda^{2}$ is 0.1 for bottomonium and 0.3 for charmonium). For moderate and high transverse momentum $P_{T} \gtrsim 2m_{Q}$ established and most successful theory that describes quarkonium production and decays in non-relativistic QCD~\cite{34,38}, and this theory is very useful for showing the accurate description of this kind of purpose.\\
In recent years, different phenomenological approaches have been proposed to describe the modification of the production cross-section of moderate and high transverse momentum quarkonia. Theoretical guidance on the relative significance of the various nuclear effects in the currently accessible transverse momentum range can be very useful.\\
Finally, we have concluded that the probability of finding the particle (charmonium $m_{J/\psi}$=1.5 GeV and bottomonium $m_{\Upsilon}$=4.5 GeV) depends upon the scattering angle ($\theta^o$), temperature ($T/T_c$) and baryonic chemical potential $\mu_{b}$. The present results indicates that the baryonic chemical potential show small effect in comparison to the temperature. Further, the differential scattering cross-section decreases rapidly with increases in impact parameter as well as the transverse momentum. This work might be helpful in understanding the process of quarkonia production under different parameters such as temperature, baryonic chemical potential etc. It is also useful to investigate the scattering rates of quarkonia. It would also provide a large amount of information regarding the internal structure of the colliding particles.\\ 

\section*{Data Availability}
This is a theoritical work and all previous results are listed in the references.\\

\section*{Conflicts of Interest}
The authors declare no competing interest.\\

\section*{Acknowledgement}
One of the author, VKA acknowledge the Science and Engineering research Board (SERB) Project No. {\bf EEQ/2018/000181} New Delhi for providing the financial support. We record our sincere gratitude to the people of India for their generous support for the research in basic sciences.

\end{document}